# A Fast local Reconstruction algorithm by selective backprojection for Low-Dose in Dental Computed Tomography*

Yan Bin(闫镔)[1;1)], Deng Lin(邓林)[1;2)], Han Yu(韩玉)[1], Zhang Feng(张峰)[1], Wang Xian Chao(汪先超)[1] and Li Lei (李磊)[1;3)]

[1] National Digital Switching System Engineering and Technological Research Center, Zhengzhou 450002, China

**Abstract**

High radiation dose in computed tomography (CT) scans increases the lifetime risk of cancer, which become a major clinical concern. The backprojection-filtration (BPF) algorithm could reduce radiation dose by reconstructing images from truncated data in a short scan. In dental CT, it could reduce radiation dose for the teeth by using the projection acquired in a short scan, and could avoid irradiation to other part by using truncated projection. However, the limit of integration for backprojection varies per PI-line, resulting in low calculation efficiency and poor parallel performance. Recently, a tent BPF (T-BPF) has been proposed to improve calculation efficiency by rearranging projection. However, the memory-consuming data rebinning process is included. Accordingly, the chose-BPF (C-BPF) algorithm is proposed in this paper. In this algorithm, the derivative of projection is backprojected to the points whose x coordinate is less than that of the source focal spot to obtain the differentiated backprojection (DBP). The finite Hilbert inverse is then applied to each PI-line segment. C-BPF avoids the influence of the variable limit of integration by selective backprojection without additional time cost or memory cost. The simulation experiment and the real experiment demonstrated the higher reconstruction efficiency of C-BPF.

## 1. Introduction

*Project supported by the National High Technology Research and Development Program of China (Grant No. 2012AA011603) and the National Natural Science Foundation of China (61372172)

1) E-mail:tom.yan@gmail.com
2) E-mail:lin.deng2130@gmail.com
3) E-mail:leehotline@gmail.com

X-ray computed tomography (CT) has been widely used in dental CT for diagnosis. One of the objectives of modern CT is to reduce radiation dose because of the health risks caused by X-rays. In dental CT, projection data is collected by irradiating the brain. The global reconstruction algorithm requires projection without truncated data to obtain the image of dens, which means the brain must be irradiated by higher radiation dose. However, the local reconstruction algorithm can deal with truncated projection data[1].

Recently inspired by ATRACT algorithm[2] which can restrain the truncation artifacts to some extent by using local filter and global filter, Wang[3] proposed an FBP algorithm based on the Radon inversion transform (RIT) and achieved fast reconstructing without reducing the quality of reconstruction. But the ability to restrain the truncation artifacts is limited.

The backprojection-filtration (BPF) algorithm can be used for local reconstruction. This algorithm was developed for image reconstruction on PI-line segments in a helical cone-beam scan, and it provided a strategy for reconstructing the exact region of interest (ROI) using truncated data[4]. Because the circular scanning trajectory is easy to implement and control in practice, it is widely used in CT[11]. Based on the concept of virtual PI-line and virtual circular orbit, Yu et al. modified the BPF algorithm to reconstruct images in a circular cone-beam scan[12]. The modified algorithm is an excellent work for low-dose dental CT because it can deal with the local reconstruction in the short scan. However, the modified algorithm is not a good choice for practice dental CT because of its low reconstruction efficiency and poor parallel performance[13]. The tent backprojection-filtration (T-BPF) algorithm[13] has been recently developed to obtain fast reconstruction from truncated data in circular cone-beam CT. This algorithm also improves reconstruction efficiency and parallel performance by make projection data rebinned into tent-like parallel-beam format from cone-beam format. And it has the same parallel performance as the FDK algorithm[14]. However, data rebinning is included in the T-BPF algorithm. Data rebinning not only introduces errors caused by trilinear-interpolation but also uses a considerable amount of memory to store the projection data acquired by the detector at all views and the rebinned projection data.

In this work, C-BPF was developed for fast reconstruction from truncated data in circular cone-beam CT. In C-BPF, the derivative of the projections is firstly backprojected to the points whose x coordinate is less than that of the source focal spot to obtain the differentiated backprojection (DBP). The finite Hilbert inverse[15] is then applied to each PI-line segment to reconstruct images from the DBP. Compared with the T-BPF algorithm, C-BPF avoids the influence of the variable integration interval by selective backprojection instead of data rebinning. Only one view projection is loaded at any time in C-BPF. Therefore, the reconstruction efficiency of C-BPF is considerably higher than that of the BPF algorithm. The memory cost of C-BPF is also considerably less than that of the BPF and T-BPF algorithms.

Similar to the BPF and T-BPF algorithms, C-BPF can also use the projection data acquired in a short scan, which can reduce the dose of radiation and the time of data acquisition under the same sampling frequency, compared with algorithms in a full scan. C-BPF can also deal with the truncated projection data, which can reduce the dose of radiation and achieve local reconstruction, compared with the global reconstruction algorithm.

This paper is organized as follows. Section II describes the geometrical structure, recalls the BPF and T-BPF algorithms, and introduces the C-BPF algorithm in detail. Section III demonstrates the advantages of C-BPF based on the results of reconstruction in the simulation experiment and the real experiment by comparing with the original BPF algorithm and the T-BPF algorithm. Finally, Section IV gives the conclusion.

## 2. Methods

This section describes the geometrical structure for circular cone-beam scan, recalls the BPF and T-BPF algorithms, and describes C-BPF in detail.

**2.1 Geometrical Structure**

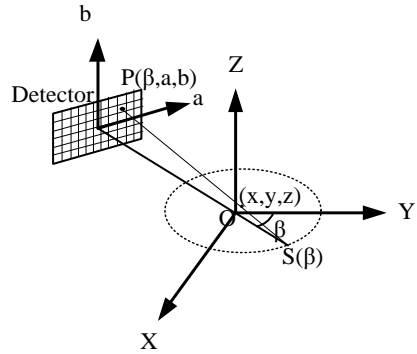

**Fig. 1. Geometrical structure for circular cone-beam scan.**

The cone-beam geometry is shown in Fig. 1. $Oxyz$ represents a Cartesian coordinate system of world. $Z$ is the rotation axis of scanner. $S(\beta)$ is the position of the source focal spot in the $Oxz$ plane and uniquely determined by the angular parameter $\beta$. The position of the source focal spot can be expressed using the following equation:

$$S(\beta) = R \cdot (\cos\beta, 0, \sin\beta), \beta \in [\beta_{start}, \beta_{end}] \quad (1)$$

where R is the distance between the origin of the Cartesian coordinate system of world and the source, and $\beta_{start}$, $\beta_{end}$ correspond to the starting and ending points of the circular orbit respectively. Projection data are collected on a flat-panel detector. $(a,b)$ represents the local Cartesian coordinate system of detector. The distance between the source and the detector is fixed, and the origin of the Cartesian coordinate system of detector is at the line which is determined by the source and the origin of the Cartesian coordinate system of world. Every X-ray and its projection can be uniquely determined by $P(\beta, a, b)$.

**2.2 BPF and T-BPF**

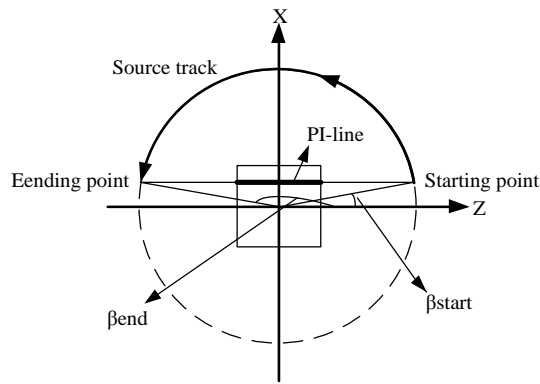

**Fig. 2. Relation between PI-line and scan angle in BPF.**

The DBP is described, which is the first of the two steps in the BPF algorithm[10] and it can be obtained by backprojecting the derivative of the projections, as shown in Formula (2):

$$b(\vec{r}) = \int_{\beta_{start}}^{\beta_{end}} P'(\beta, a_r, b_r) d\beta \tag{2}$$

where $r(x, y, z)$ denotes any point on a PI-line, $P'(\phi, a_r, b_r) = \partial P(\phi, a_r, b_r) / \partial a_r$, and $[\beta_{start}, \beta_{end}]$ denotes the view-angle integration interval of backprojection. In general, we make every PI-line parallel to Z axis, and every PI-line can be uniquely determined by $PI(x, y)$ (the same applies to T-BPF and C-BPF). Therefore, the integration interval of backprojection is the same for PI-line segments on the plane x=k, where k is an arbitrary constant. The relationship between PI-line and scan angle is shown in Fig. **2**, and $\beta_{start}$ and $\beta_{end}$ are determined by the plane x=k, as shown in Formula (3):

$$\begin{cases} \beta_{start} = \arcsin(x/R) \\ \beta_{end} = \pi - \arcsin(x/R) \end{cases} \tag{3}$$

The object function $f(\vec{r})$ and $b(\vec{r})$ have the following relation[10][16]:

$$b(\vec{r}) = -2\pi H f(\vec{r}) \tag{4}$$

where $Hf$ presents the Hilbert transform of $f$. The object function $f(\vec{r})$ can be obtained by applying the inverse of finite Hilbert transform[15]. The BPF algorithm for CBCT is performed as follows:

1) Load all projections and determine the derivative of the projection.

2) Select a plane x=k for reconstruction and determine the accordant $\beta_{start}$ and $\beta_{end}$.

3) To obtain the DBP of the plane x=k, backproject the derivative of the projection that corresponds to $\beta \in [\beta_{start}, \beta_{end}]$ on the points that are in the plane.

$$b(\vec{r}) = \int_{\beta_{start}}^{\beta_{end}} P'(\beta, a_r, b_r) d\beta, \vec{r} \in \{(x, y, z) | x = k\} \tag{5}$$

4) Perform the finite Hilbert inverse on those PI-line in the plane *x=k*. Repeat step 2 if not all planes are backprojected.

Analysis for time cost of the BPF algorithm: Four circles are present in its backprojection, in which three are dimensional direction circles and one is a view-angle circle. The integration limit of view-angle is determined by the location of the plane *x=k*. Thus, a dimensional direction circle, must be implemented before the view-angle circle, which produces more calculations of trigonometric function and the intermediate variable that corresponds to the y coordinate or z coordinate in backprojection. The above operations account for the low reconstruction efficiency of the BPF algorithm. In addition, the relativity of the circles accounts for the low parallel performance of backprojection. Analysis for memory cost of the BPF algorithm: The limit of integration for backprojection (about $\pi$) is required by every plane. Therefore, all projections must be loaded in the memory before; otherwise, every projection will be read repeatedly from the hard disk[20].

To improve parallel performance and reconstruction efficiency, T-BPF is performed by firstly rebinning the cone-beam data to tent-like parallel-beam data and then applying the BPF-type algorithm to reconstruct images.

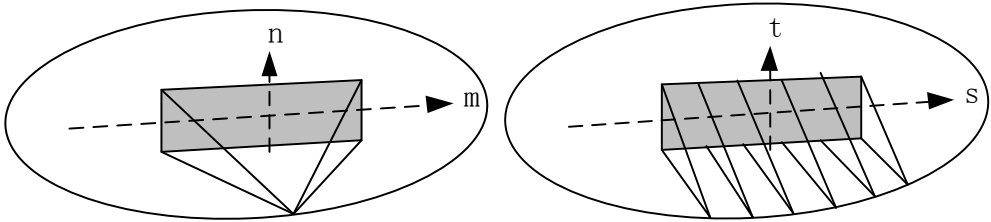

**Fig. 3. Cone-beam projection data at a view (left) tent-like parallel projection data at a view after rebinning (right), where** $(t,s)$ **are the coordinates on virtual detector in T-BPF.**

A virtual detector is introduced to describe the formula of rebinning conveniently. The virtual detector is placed on the center of rotation and parallel to the real detector. $P(\beta,m,n)$ denotes the projection data acquired by the virtual detector and $(m,n)$ is the position on the virtual detector; The rebinned projection data can be parameterized by $P_{tent}(\theta,s,t)$ and $(s,t)$ is the position After rebinning. The rebinning formula is implemented using the following equation:

$$P_{tent}(\theta,s,t) = P\left(\theta - \arcsin\frac{s}{R}, \frac{sR}{\sqrt{R^2-s^2}}, \frac{tR^2}{R^2-s^2}\right) \tag{6}$$

After rebinning, the cone-beam data are arranged into parallel-beam data on a virtual rectangular detector plane shown in Fig. 3. In the image reconstruction from the parallel-beam projection data, the integration limit of view-angle for backprojection is fixed (from 0 to $\pi$) for every PI-line, and the DBP can be described as

$$b(\vec{r}) = \int_0^\pi P'_{tent}(\theta, a_r, b_r) d\theta \tag{7}$$

The T-BPF algorithm for CBCT is performed as follows:

1) Load all projections and determine the derivative of the projection.
2) Rearrange the cone-beam data to tent-like parallel-beam data.
3) To obtain the DBP of the $f(x)$, backproject the derivative of the projection that corresponds to $\beta \in [0, \pi]$ to $f(x)$.
4) Carry out the finite Hilbert inverse on the PI-line.

Analysis for time cost of the T-BPF algorithm: The T-BPF algorithm avoids the influence of the variable limit of integration for the backprojection by projection data rebinning. Thus, the T-BPF algorithm has higher reconstruction efficiency than the BPF algorithm. Analysis for memory cost of the T-BPF algorithm: compared with the BPF algorithm, additional memory is used to keep the rebinned projection and the memory cost for the T-BPF algorithm is about twice over that for the BPF algorithm.

**3.3 C-BPF**

Backprojection is the most time-consuming part in the BPF algorithm. T-BPF improves backprojecting efficiency by rearranging projection data because the variable limit of integration of backprojectin is converted to fixed limit of integration. However, T-BPF introduces the data rebinning process. In this paper, we proposed C-BPF, which improves backprojecting efficiency by selective backprojection without rearranging projection data, reading projection repeatedly from hard disk, or loading all projections beforehand. C-BPF has greatly reduced memory cost compared with the BPF and T-BPF algorithms. C-BPF also greatly reduces time cost compared with the BPF algorithm.

We modified Formula (7) to avoid the influence of the variable limit of integration for the backprojection.

Formula (7) can be described in detail using the following equation:

$$b(x, y, z) = \int_{X_{min}}^{X_{max}} \delta(\tilde{x} - x) \cdot \int_{\beta_{start}}^{\beta_{end}} \int_{Y_{min}}^{Y_{max}} \delta(\tilde{y} - y) \int_{Z_{min}}^{Z_{max}} \delta(\tilde{z} - z) P'(\beta, a_r, b_r) d\tilde{z} d\tilde{y} d\beta d\tilde{x} \tag{8}$$

where $\delta(x)$ is pulse function; $X_{min}$, $X_{max}$, $Y_{min}$, $Y_{max}$, $Z_{min}$, and $Z_{max}$ denote the minimum and maximum of objects in the x, y, and z directions, respectively. To make the limit of integration the same for every PI-line, Formula (8) is transformed as follows:

$$b(x,y,z) = \int_{X_{\min}}^{X_{\max}} \delta(\tilde{x}-x) \cdot \int_0^{2\pi} u(\beta-\beta_{start})u(\beta_{end}-\beta) \int_{Y_{\min}}^{Y_{\max}} \delta(\tilde{y}-y) \int_{Z_{\min}}^{Z_{\max}} \delta(\tilde{z}-z)\mathrm{P}'(\beta,a_r,\mathrm{b}_r)d\tilde{z}d\tilde{y}d\beta d\tilde{x}$$

(9)

where $u(x)$ is the step function. In this way, integrating factors $dx$, $dy$, $dz$, and $d\beta$ are unattached. Thus, the order of integration can be changed as

$$b(x,y,z) = \int_0^{2\pi} \int_{X_{\min}}^{X_{\max}} \int_{Y_{\min}}^{Y_{\max}} \int_{Z_{\min}}^{Z_{\max}} u(\beta-\beta_{start})u(\beta_{end}-\beta)\delta(\tilde{x}-x)\delta(\tilde{y}-y)\delta(\tilde{z}-z)\mathrm{P}'(\beta,a_r,\mathrm{b}_r)d\tilde{z}d\tilde{y}d\tilde{x}d\beta$$

(10)

$x < R\sin\beta$ can be obtained from the geometry relation between the angle $\beta_{start}$ $\beta_{end}$ and PI-line Formula (3). Thus, Formula (10) can be simplified as follows:

$$b(x,y,z) = \int_0^{2\pi} \int_{X_{\min}}^{X_{\max}} \int_{Y_{\min}}^{Y_{\max}} \int_{Z_{\min}}^{Z_{\max}} u(\mathrm{R}\sin\beta-x)\delta(\tilde{y}-y)\delta(\tilde{z}-z)\mathrm{P}'(\phi,a_r,\mathrm{b}_r)d\tilde{z}d\tilde{y}d\tilde{x}d\beta \quad (11)$$

No point is backprojected when the source is at the broken line in Fig. 4. Thus, the projection obtained when the source is at the broken line does not contribute to reconstruction, so that the limit of integration of β can be cut short and Formula (11) can be written as follows:

$$b(x,y,z) = \int_\psi^{\pi-\psi} \int_{X_{\min}}^{X_{\max}} \int_{Y_{\min}}^{Y_{\max}} \int_{Z_{\min}}^{Z_{\max}} u(\mathrm{R}\sin\beta-x)\delta(\tilde{y}-y)\delta(\tilde{z}-z)\mathrm{P}'(\phi,a_r,\mathrm{b}_r)d\tilde{z}d\tilde{y}d\tilde{x}d\beta \quad (12)$$

where $\psi = \arcsin(X_{\min}/R)$. Compared with the BPF algorithm, C-BPF is performed in such a way that plane x=k is selected by the projection of every view instead of the way the projection is selected by plane x=k. The shadow in Fig. 4 is backprojected by the derivative of the projection that corresponds to $\beta$. Only one projection is loaded at any time, and every projection is loaded only one time in memory. After backprojecting the derivative of the projection that corresponds to $\beta \in [\psi, \pi-\psi]$ (where $\psi = \arcsin(X_{\min}/R)$), every PI-line is backprojected by the derivative of the projection that corresponds to $\beta \in [\beta_{start}, \beta_{end}]$, and the DBP of $f(x)$ is obtained. The C-BPF algorithm for CBCT is performed as follows:

1) Determine the $\psi$.
2) Load a projection ($\beta \in [-\psi, \pi+\psi]$), and determine the derivative of the projection.

3) Backproject the derivative of the projection to the PI-line, where $x < R\sin\beta$.

4) Proceed to step 5) if all projections ($\beta \in [-\psi, \pi+\psi]$) are used or return to step 2).

5) Carry out the finite Hilbert inverse on the PI-line.

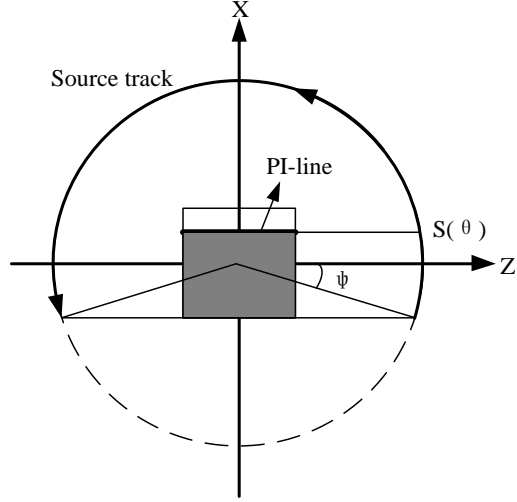

**Fig. 4. Relationship between PI-line and scan angle in C-BPF.**

## 3. Implementation and Experiment

To testify the advantages of the C-BPF method presented above, the three algorithms were implemented in C language for the simulation experiment and the real experiment. The experiments were performed by using the CPU(Inter(R) Xeon(R) X5450 @3.00GHz).

### 3.1 Simulation experiment

The standard 3D Shepp-Logan head phantom[18] was used to be reconstructed to compare C-BPF with the T-BPF and BPF algorithms. The length, width, and height of the head phantom were 7.14331 mm. The distance between the source and the rotation axis of scanner was 477mm and the distance between the source and the detector was 1265mm. The cone-beam projection data were acquired from a flat-panel detector with 500 pixels×500 pixels and 0.148mm per pixel. 360 projection views were uniformly distributed over the 2π circular trajectory. However, all the algorithms only used the views in the range of π plus twice the cone angle. The final reconstruction was done on a 256×256×256 grid.

The root mean square error (RMSE)[19] was introduced to evaluate the reconstruction results, and it can be written as follows:

$$RMSE = \sqrt{\frac{1}{N}\sum_{i=1}^{N}[f_t(i) - f_0(i)]^2} \qquad (13)$$

where $f_t$ and $f_0$ denote the reconstructed and the reference images of voxels.

Fig. 5 presents the model of Shepp–Logan head phantom and the reconstruction results of three algorithms for the model. The three orthogonal planes ($x=0$, $y=0$ and $z=0$) of 3D images reconstructed by the BPF, T-BPF and C-BPF algorithms were compared. At the same time, the corresponding profiles on the middle horizontal line are shown in Fig. 6.

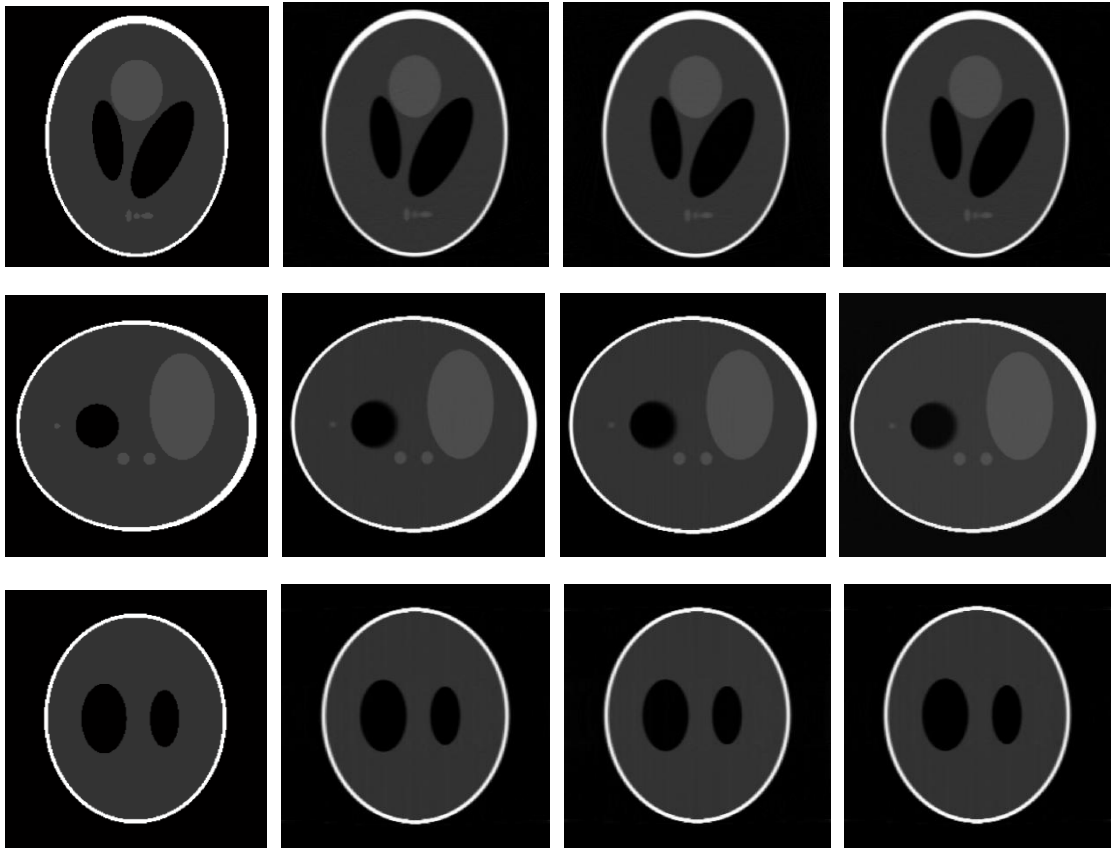

**Fig. 5. 2D slices in 3D images reconstructed using the BPF, T-BPF, and C-BPF algorithms. The 128th slices of the plane z=0, plane x=0 and plane y=0 are represented in the first, second, and third rows respectively. The slices of the Shepp-Logan model and the reconstruction results of the BPF, T-BPF, and C-BPF algorithms are represented in the first, second, third, and fourth columns respectively.**

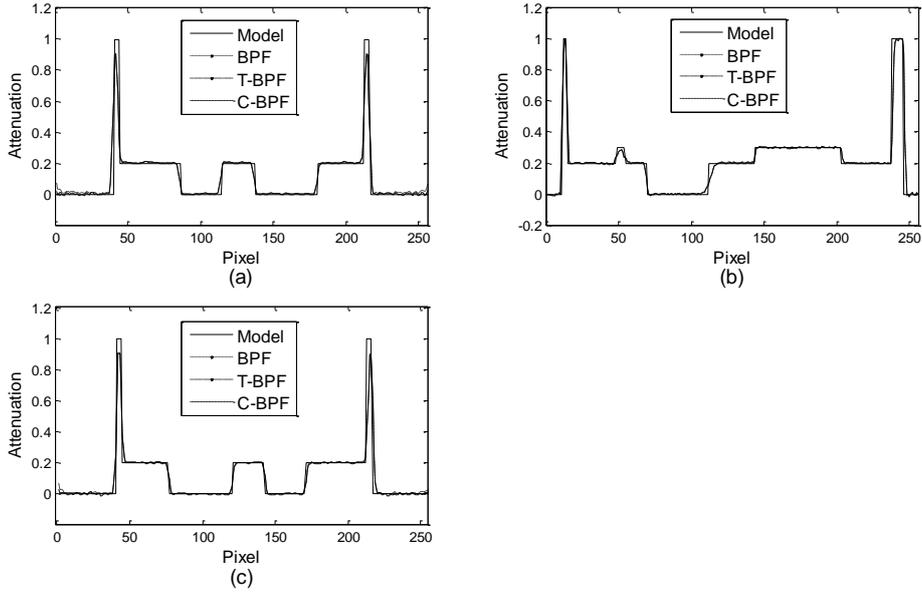

**Fig. 6. Profiles of the images shown in Fig. 5 along the horizontal middle lines. (a), (b), and (c) are the profiles of the first, second, and third rows of Fig. 5, respectively.**

**Table 1. Reconstruction results of BPF, T-BPF, and C-BPF algorithms.**

| ALGORITHMS | SIZE | Time (minute) | Memory (Mb) | RMSE |
|---|---|---|---|---|
| BPF | 256×256×256 | 16.05 | 417.36 | 0.0708 |
| T-BPF | 256×256×256 | 12.18 | 753.51 | 0.0701 |
| C-BPF | 256×256×256 | 11.71 | 49.23 | 0.0708 |

respectively.

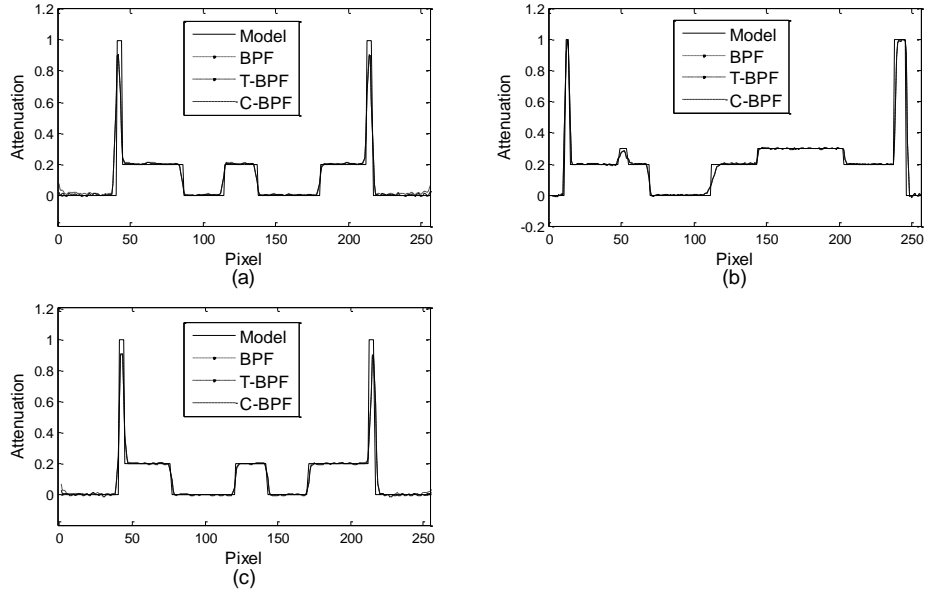

Fig. 6 and Table 1 show no significant differences among reconstruction results of the original BPF algorithm, the T-BPF algorithm and C-BPF. However, in the case of serial computing, C-BPF is more efficient and memory saving than the BPF and T-BPF algorithms because that the backprojection implementation of C-BPF not only reduced many repetitive calculations of trigonometric function, but also invoked no additional intermediate variables that corresponded to the y coordinate or z coordinate in backprojection without projection data rebinning. The memory required by C-BPF is considerably less than that required by the BPF and T-BPF algorithms because all projections for the original BPF and T-BPF algorithms must be loaded to eliminate repeated reading of projection data[20]. Only one view of the projection data is loaded for the C-BPF algorithm at any time.

Contrast experiments of parallel performance for the three algorithms are performed on the Tesla C1060 GPU in the environment of CUDA 4.4 runtime API.

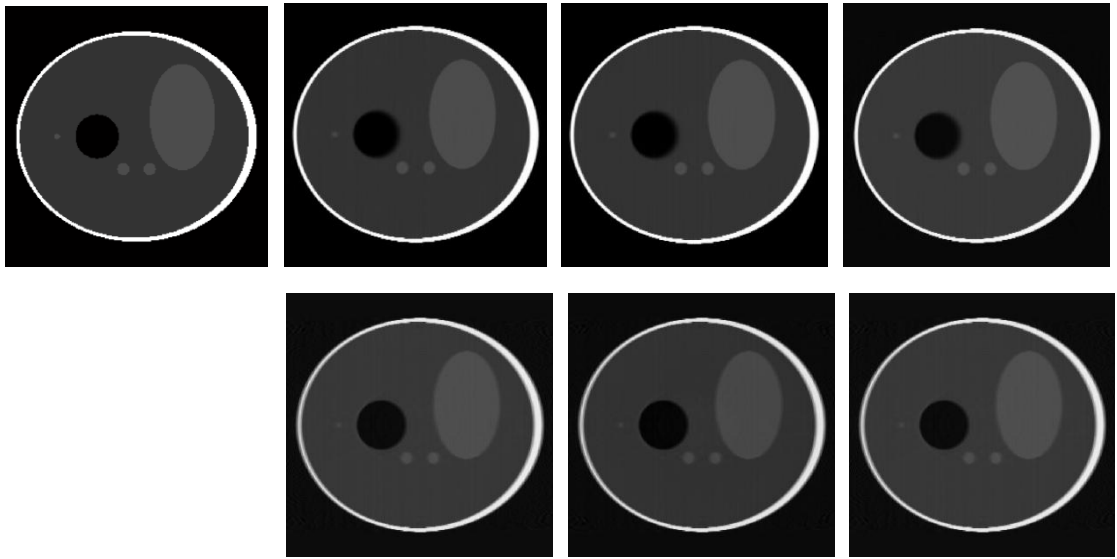

**Fig. 7. Reconstructed slice of the three algorithms by parallel computing(GPU). the Shepp-Logan model and the reconstructed slices of the BPF, T-BPF, and C-BPF algorithms are represented in the first, second, third, and fourth columns, respectively. On the right, the reconstructed slices using CPU and GPU are represented in the first and second rows, respectively.**

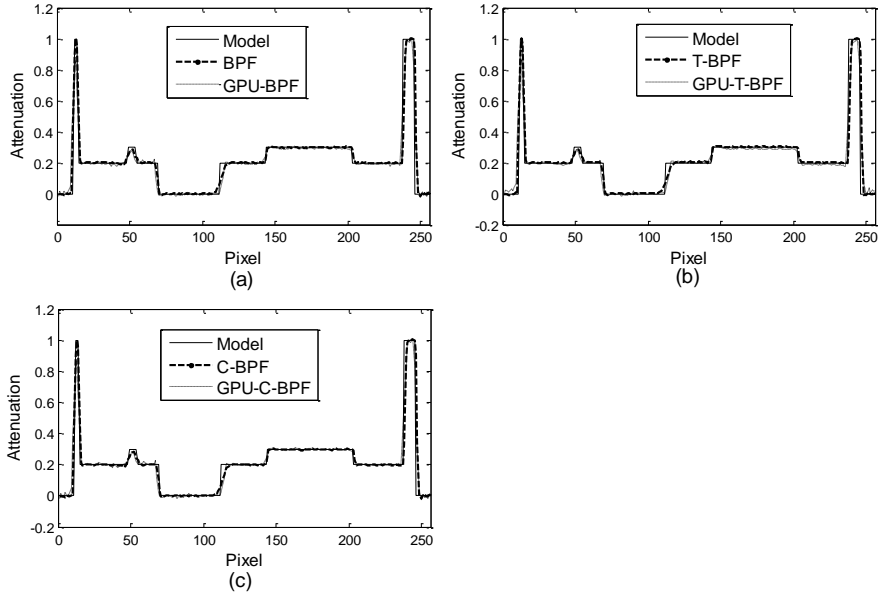

Fig. 8. Profiles of the reconstructed slices along the horizontal middle lines. (a), (b), and (c) represent the profiles of the slices reconstructed using the BPF, T-BPF, and C-BP algorithms, respectively.

Table 2. Contrast experiments of CPU and GPU.

| ALGORITHMS | SIZE | Reconstruction time cost (CPU) | Reconstruction time cost (GPU) | Speedup factor |
|---|---|---|---|---|
| BPF | 256×256×256 | 963.7 s | 3.245 s | 297 |
| T-BPF | 256×256×256 | 730.8 s | 0.705 s | 1036 |
| C-BPF | 256×256×256 | 705.6 s | 0.677 s | 1043 |

No difference can be observed between images reconstructed by CPU and GPU (Figs. 7 and 8). As shown in Table **2**, the time costs for parallel performance by the BPF, T-BPF and C-BPF algorithm are 3.245, 0.705, and 0.677s respectively. The parallel performance of the BPF algorithm is poor because of relations between a spatial direction circle and the scanning angle circle in the backprojection. The parallel performance of the T-BPF algorithm is better than that of the BPF algorithm because the four circles is independent in backprojection implementation. The parallel performance of C-BPF is slightly better than that of the T-BPF algorithm because no data rebinning is included and the relations between angle circle and a 8spatial direction circle is eliminated by neglecting some PI-lines in backprojection implementation.

### 3.2 Real experiment

Reconstructions of real data were performed for three algorithms and the real data was acquired with the cone-beam CT system which was mainly consisted of the flat-panel detector (Varian4030E, USA) with a pixel size of 0.127mm, the object holder(which can be circumvolved) and the X-ray source (Hawkeye 130, Thales, France).

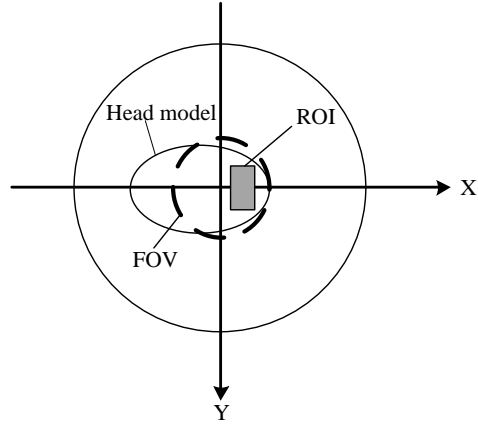

**Fig. 9 . the Planform of ROI.**

Experiments were carried out to reconstruct an ROI of the head model and check the capability of the C-BPF method (Fig. 9). The distance between source and the rotation axis of scanner was 678 mm and the distance between source and the detector was 1610 mm. The projection data was acquired from the flat-panel detector with 3200 pixels × 2304 pixels and 0.128 mm per pixel. 360 projection views were uniformly distributed over the $2\pi$ circular trajectory. The size of the reconstructed ROI was 700 pixels × 1200 pixels × 500 pixels with 0.107 mm per pixel. The projection view used by the three algorithms varied from $\Psi$ to 180-$\Psi$. The distance between the rotation axis of scanner and the ROI center was 450 pixels(48.15mm). Thus, the following equation applies:

$$\psi = \arcsin\left(\frac{(450-700/2)\times 0.107}{678}\right)\times \frac{180}{\pi} = 9.080^{o} \qquad (14)$$

The data acquired on the detector at any view angle were not used completely. We only used the middle 1600 pixels in $b$ axe direction. The data type of the projection data was changed to unsigned short from float.

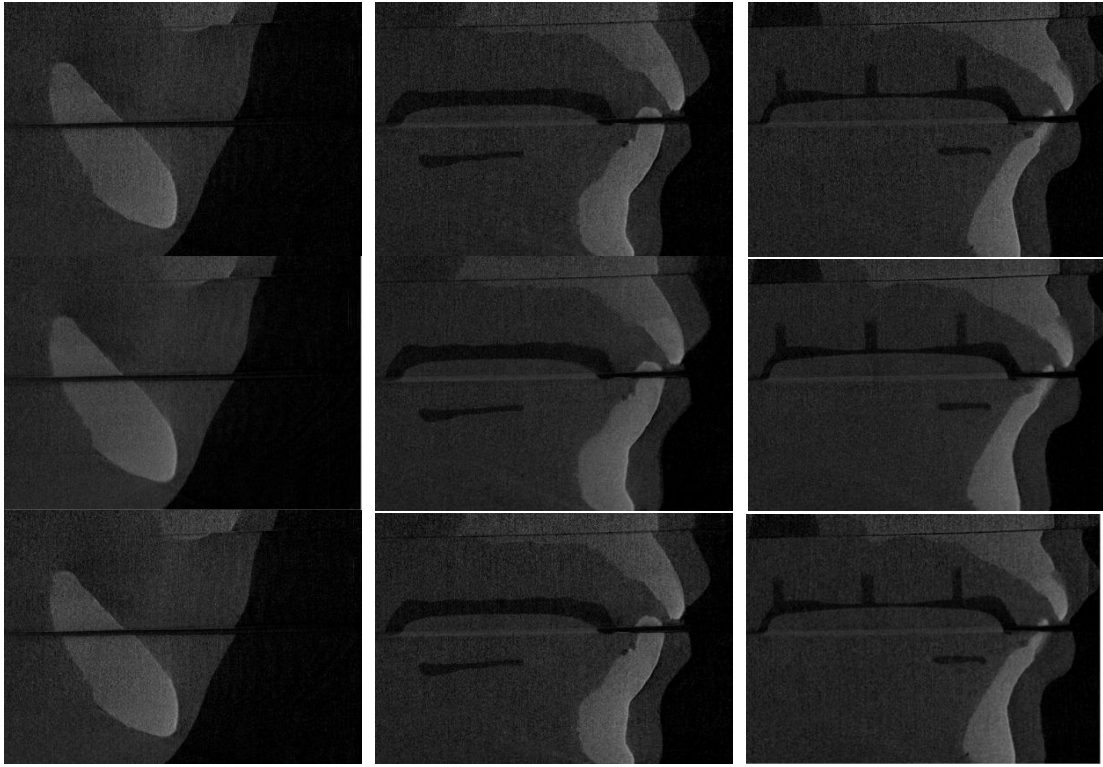

**Fig. 10. 2D slices in 3D images reconstructed using the BPF, T-BPF, and C-BPF algorithms. The slices of the plane y=26.86 mm, plane y= 54.85 mm and plane y=61.83 mm are represented in the first, second, and third rows respectively. The reconstruction results of the BPF, T-BPF, and C-BPF algorithms are represented in the first, second and third columns respectively.**

**Table 3. Reconstruction results of BPF, T-BPF, and C-BPF algorithms**

| ALGORITHMS | SIZE | Time (minutes) | Memory (GB) |
|---|---|---|---|
| BPF | 700×500×1200 | 668.54 | 8.54 |
| T-BPF | 700×500×1200 | 249.73 | 15.77 |
| C-BPF | 700×500×1200 | 213.38 | 1.67 |

Fig. 10 shows the correctness of C-BPF again by reconstructing the ROI from the truncated data. As shown in Fig. 10 and Table 3, the three algorithms with truncated data result in reconstructed images with no significant differences, and the advantages on memory cost and time cost of C-BPF are evident. In the case of comparable reconstruction quality, the reconstruction efficiency of C-BPF is the highest, and its memory costing is the least among the three algorithms. In practice, the memory cost of an algorithm counts. If the memory cost is too high to be satisfied, some strategies must be introduced, such as lowering the spatial resolution of the reconstruction images or decomposing the backprojection by the limiting of integration of spatial direction. However, the two resolution results have lower reconstruction quality or greater time cost.

Large scale reconstruction is needed to reconstruct the high spatial resolution or large size of ROI. To achieve large scale reconstruction, BPF and T-BPF require considerable amounts of memory, which cannot be satisfied because of limited hardware. For the BPF algorithm, reading projection from the hard disk repeatedly or block-dividing backprojection must be carried out. Block-dividing backprojection also involves the reading projection from the hard disk repeatedly. Therefore, the BPF algorithm must utilize considerably higher time cost to avoid the high memory cost in the case of large scale reconstruction. For the T-BPF algorithm, reading and writing projection on the hard disk must be carried out repeatedly. The T-BPF algorithm must use block-dividing data-rebinning (the block is divided by the index of pixel on the detector). Block-dividing data-rebinning involves many trigonometric functions or the use of some memory to store the results of the trigonometric functions that are computed beforehand. The rebinned projection must be stored on hard disk after data rebinning on a block and be read for backprojection. The required processes will cost longer time to obtain the DBP. In a word, the T-BPF algorithm must also use a considerably higher time cost to avoid the large memory cost in the case of large scale reconstruction. However, no additional process for decomposing the backprojection is needed for the C-BPF algorithm because of the low memory cost for C-BPF. Thus, C-BPF can reconstruct considerably faster than the BPF and T-BPF algorithms in the case of large scale reconstruction.

## 4. Conclusion

In the work, C-BPF was developed for image reconstruction from the truncated data acquired in a short scan. In C-BPF, the derivative of projections is backprojected to the points whose x coordinate is less than that of the source focal spot to obtain the DBP. Then, the finite Hilbert inverse is applied on each PI-line segment to reconstruct images from the DBP. Compared with the BPF and T-BPF algorithms, C-BPF has higher reconstruction efficiency, better parallel performance, and lower memory cost. The C-BPF algorithm avoids the influence of the variable limit integration by selective backprojecting without projection data rebinning and the need to load all projection data in advance. The C-BPF has achieved fast local reconstruction from the truncated data acquired in a short scan. The method of obtaining DBP can also be used for DBP-POCS[21] and SVD-THT[23]. The approach can also be applied to practical CT systems and achieve a fast local reconstruction.